# Study of Sublevel Population Mixing Effects in Hydrogen Neutral Beams

S. V. Polosatkin,[1,2,3] A. A. Ivanov,[1,2] A. A. Listopad,[1] and I. V. Shikhovtsev[1,2]

[1]*Budker Institute of Nuclear Physics SB RAS, Novosibirsk, Russia*
[2]*Novosibirsk State University, Novosibirsk, Russia*
[2]*Novosibirsk State Technical University, Novosibirsk, Russia*
e-mail: s.v.polosatkin@inp.nsk.su

Intensities of hydrogen Balmer-series lines are sensitive to distribution of populations of sublevels of excited state due to difference of radiative decay branching ratio for sublevels with different orbital quantum number. Since these sublevels are degenerate transitions between sublevels and mixing of its population may be caused by very weak action of environment. A beam of fast hydrogen atoms is a system in which these fine effects of population mixing can be observed.

Experimental data of $H_\alpha$ line radiation intensities in hydrogen neutral beams are compared with calculations assumed different models of sublevel population. The comparison most probably point out that population mixing effects cause transitions from long-lived 3s to 3p sublevel with corresponding changes in $H_\alpha$ radiation intensities. Discussed effects may influence to results of Doppler-shift spectroscopy measurements of the neutral beam parameters such as species and impurities content and neutralizing efficiency.

## INTRODUCTION

Well-known feature of energy level structure of hydrogen atom is degeneration of sublevels in orbital quantum number. Since a splitting of sublevels due to spin-orbital interaction and relativistic effects is in order of $10^{-5}$ eV only, sublevel population distribution may be sensitive to very weak action of environment. In turn, this distribution affect to a rate of radiative de-excitation of the level, and that is intensity of line radiation. Consequently, in some cases intensity of radiation of hydrogen spectral lines is determined by fine effects of mixing of sublevel populations (so-called l-mixing).

In particular, effect of sublevel population mixing can be manifested in the measurements of intensities of Doppler-shifted hydrogen Balmer-$\alpha$ lines in the beams of fast hydrogen atoms. Such beams are widely used for plasma heating and diagnostics in nuclear fusion oriented experiments. As a rule, such beams contain species with fractional energies (E/2, E/3, E/18) originated from acceleration and follows dissociation of hydrogen-contained molecular ions ($H_2^+$, $H_3^+$, $H_2O^+$). Routine method for controlling of species content is Doppler-shift spectroscopy consisting of measurements of intensities of Doppler-shifted $H\alpha$ spectral lines appeared in collisions of beam particles with background gas [1-4].

Interpretation of Doppler-shift spectroscopy measurement of neutral beams species content requires an assumption about distribution of populations of thin structure of excited state (hydrogen n=3). The reason is that in contrast to long-lived 3s and 3d sublevels that decays with $H_\alpha$ emission, the 3p sublevel predominantly decays directly to ground state with emission of $L_\beta$ line. Transitions between sublevels would change effective branching factor that is ratio of de-excitation via $H_\alpha$ and $L_\beta$ emission.

Several fine processes in the beamline may cause mixing of sublevel population and accordingly emissivity of H$_\alpha$ radiation from beam atoms. Mentioned processes are collisions of fast atoms with background plasma particles or polar molecules of residual gas (e.g. water), as well as motional Stark effect in a stray magnetic field. In the ref.[5,6] the influence of these processes is evaluated and correction factors for Doppler-shift measurements of beam species content are calculated for different levels of l-mixing.

The calculations gives the variation up to 30% in the correction factors for half- and third- energy atoms for different models of sublevel population. Most notable effect that can lead to two-fold overestimation of impurity concentration is found for correction factors for E/18 species.

Another important effect of l-mixing is a change of the ratio of radiation intensities in pure neutral and composite (that means contained both atoms and ions) beam. This ratio sometimes uses for controlling of efficiency of neutralization of primary ion beam. In this case accurate accounting of sublevel population mixing effects is vital for correct measurements.

A method for investigation of population mixing extent in the beamlines of hydrogen neutral beam injectors was proposed in the ref.5. In the presented paper the idea of the method is represented and experimental data of Doppler-shift spectroscopy of the two NB injectors are analyzed in the point of view of verifying of l-mixing in their beamlines.

## I. METHOD FOR OBSERVATION OF L-MIXING

An idea of the method bases on the comparison on intensities H$_\alpha$ radiation in the pure neutral and composite (that means contained both hydrogen atoms and ions) beams. In the neutral beam injectors primary ion beam is generated by ion source and neutralized on the hydrogen gas target. The beam on the exit of the target consists on both atoms and ions with relative concentration:

$$f_a = \frac{\sigma_{CX}}{\sigma_i + \sigma_{CX}} - \frac{\sigma_{CX}}{\sigma_i + \sigma_{CX}} \exp\left[-(\sigma_i + \sigma_{CX})\cdot\overline{nl}\right]$$

$$f_i = \frac{\sigma_i}{\sigma_i + \sigma_{CX}} + \frac{\sigma_{CX}}{\sigma_i + \sigma_{CX}} \exp\left[-(\sigma_i + \sigma_{CX})\cdot\overline{nl}\right]$$

where $f_a$, $f_i$ – relative concentrations of atoms and ions on the exit of the neutralizer, <nl> - linear density of the target, and $\sigma_i$ and $\sigma_{CX}$ – cross-sections of ionization and charge exchange in the gas target. As a rule, ion fraction of the beam is separated by special magnet, and Doppler-shift spectroscopy system is placed after this separation unit.

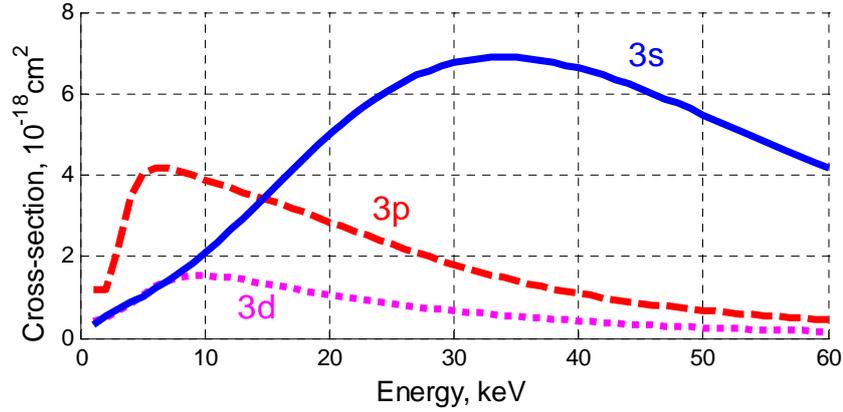

*Fig. 1. Cross-sections of charge-exchange excitation on the sublevels of n=3 level in $H^+ + H_2 \rightarrow H^* + H_2^+$ collisions. Solid line – 3s sublevel, dashed line – 3p, dotted line – 3d. Data from ORNL-6086 [6]*

Investigation of l-mixing extent can be done by comparison of $H_\alpha$ radiation in composite beam (measured with switched off magnet) and pure neutral beam (with energized separating magnet). The ratio of intensities for this case is equal to

$$R_c = \frac{I_c}{I_p} = \frac{\sigma_{ex}^0(E) \cdot B^0 \cdot f_a(E) + \sigma_{ex}^+(E) \cdot B^+ \cdot f_i(E)}{\sigma_{ex}^0(E) \cdot B^0 \cdot f_a(E)}$$

where $\sigma_{ex}^0$ and $\sigma_{ex}^+$ - cross-sections of production of exciting (n=3) atom in the collisions of atoms and ions with background gas, $B^0$, $B^+$ - effective branching ratios for $H_\alpha$ generation. The latest branching ratio is sensitive to l-mixing process. Actually, charge-exchange excitation process predominantly produces atoms on long-lived 3s sublevel (see fig.1). Since radiative transition from this sublevel to ground state forbidden by selection rules, the branching ratio should be close to unity. On the contrary, if characteristic

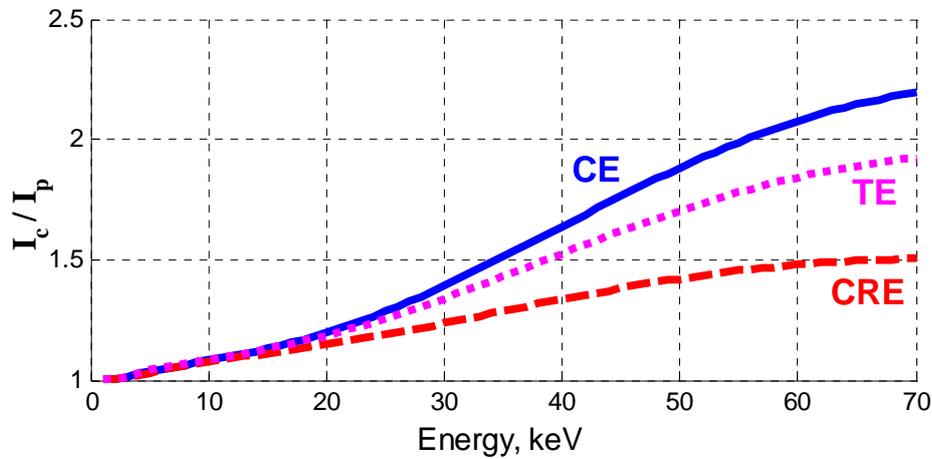

*Fig.2. The ratios of Doppler-shifted $H_\alpha$ radiation intensities for composite and pure neutral beams. Calculations for thick neutralizing target ($<nl> \rightarrow \infty$) for different models of sublevel population distribution*

time of l-mixing less then lifetime of 3s sublevel (158 ns) dominating de-excitation scheme should be transition to 3p sublevel with follows radiative decay with branching ratio 0.12.

Calculated values of R for thick neutralizing target ($<nl> \to \infty$) and three ultimate cases of excited sublevel population are shown on fig.2. These cases are specified as follows:

**Coronal equilibrium (CE)**: the case of absence of l-mixing transitions

$$\sigma_{ex} B^{CE} = \sigma_{3s} + \sigma_{3p} \frac{A_{3p-2s}}{A_{3p-2s} + A_{3p-1s}} + \sigma_{3d} = \sigma_{3s} + 0.12 \cdot \sigma_{3p} + \sigma_{3d}$$

**Thermal equilibrium (TE):** the case of strong l-mixing with thermal equilibrium distribution of sublevel populations

$$\sigma_{ex} B^{TE} = \left(\sigma_{3s} + \sigma_{3p} + \sigma_{3d}\right) \frac{g_{3s} A_{3s-2p} + g_p A_{3p-2s} + g_{3d} A_{3d-2p}}{g_{3s} A_{3s-2p} + g_{3p} A_{3p-2s} + g_{3d} A_{3d-2p} + g_{3p} A_{3p-1s}} = 0,43 \cdot \left(\sigma_{3s} + \sigma_{3p} + \sigma_{3d}\right)$$

**Collision-radiative equilibrium (CRE):** intermediate case consisted in decay of 3s sublevel via transition to 3p (with branching 0.12) and direct decay of 3d

$$\sigma_{ex} B^{CRE} = \frac{A_{3p-2s}}{A_{3p-2s} + A_{3p-1s}} \left(\sigma_{3s} + \sigma_{3p}\right) + \sigma_{3d} = 0.12 \cdot \left(\sigma_{3s} + \sigma_{3p}\right) + \sigma_{3d},$$

where $\sigma_{3s}$, $\sigma_{3p}$, $\sigma_{3d}$ – sublevel-resolved excitation cross-sections. Notice that the models specify equilibrium between sublevels of thin structure of principal states while pure coronal equilibrium over principal states is assumed. In the following section we will compare these calculated values of R with available experimental data to find out which type of equilibrium is realized in practice.

## II. DOPPLER-SHIFT MEASUREMENTS OF Hα RADIATION OF NEUTRAL BEAMS

First source of available experimental data is diagnostic neutral beam injector RUDI operating on the tokamak TEXTOR [8]. Two sets of measurements with different effective thickness on neutralizing target are presented in the ref.9. These sets of data consist on Doppler-shifted $H_\alpha$ spectra taken in the same parameters of ion source with and without ion separation. The accelerating voltage of ion source is $E_{inj}$=50 kV; since the beam includes species with fractional energies experimental values for several energies (exactly 50, 25, and 16.7 keV) can be found from every set of the data (the value for the energy $E_{inj}/18$=2.8 keV is not sensitive to model of equilibrium because of low relative concentration of ions in the beam and so is not accounted).

Second source of data is diagnostic injector for W7-X stellarator has developed in BINP. Three sets of measurements with injection energy $E_{inj}$=59 keV and slightly different regimes of gas puffing were produced in preliminary tests of the injector.

An aim of the work is to distinguish model of sublevel population most consistent with experimental data. The one problem of comparison of measurements with calculations is absence of sustainable data about neutralizing target thickness. Accordingly, likelihood functions for <nl> value in the framework of different equilibrium models were calculated and compared based on experimental data. The likelihood function is estimated as

$$L\left(\overline{nl}\right) = \prod_{i=1\div 3} \frac{1}{\sqrt{2\pi}\Delta R_i} \exp\left(-\frac{\left(R_c\left(\overline{nl}\right) - R_i\right)^2}{2\Delta R_i^2}\right)$$

where $R_c$ and $R_i$ – calculated and measured $H_\alpha$ intensity ratios for different energies of particles, $\Delta R_i$ – tolerances of the measurements. These tolerances should be evaluated from accuracy of data fitting, shot-to-shot reproducibility of the measurements and other sources of errors. Here we use background level of $H_\alpha$ spectra as upper estimation of tolerance of radiation intensity. Actually the following results of comparison of experimental data with model calculation are not sensitive to the estimations of tolerances.

Likelihood values for all sets of data calculated for different models of equilibrium are presented on the fig.3a; experimental values of R with least square fits corresponding to maximal value of likelihood are shown on the fig.3b. Notice that the curves on the fig 3b are calculated for different values of target thickness <nl> corresponded to best fits of experimental data. Consistency of the models of equilibrium with experimental data can be specified by Pearson $\chi^2$ test. The values of $\chi^2$ for all data sets are presented in the table 1:

| Set of the data | $\chi^2$ / <nl> [$10^{16}$ cm$^{-2}$] | | |
|---|---|---|---|
| | **CRE** | **CE** | **TE** |
| RUDI 1 | 0.4 / 0.15 | 4.8 / 0.31 | 3.0 / 0.25 |
| RUDI 2 | 0.55 / 0.34 | 0.35 / 0.97 | 0.37 / 0.64 |
| W-7X 1 | 1.9 / 0.13 | 7.6 / 0.29 | 5.6 / 0.22 |
| W-7X 2 | 2.0 / 0.13 | 6.3 / 0.26 | 4.5 / 0.21 |
| W-7X 3 | 0.6 / 0.14 | 4.5 / 0.28 | 3.8 / 0.22 |

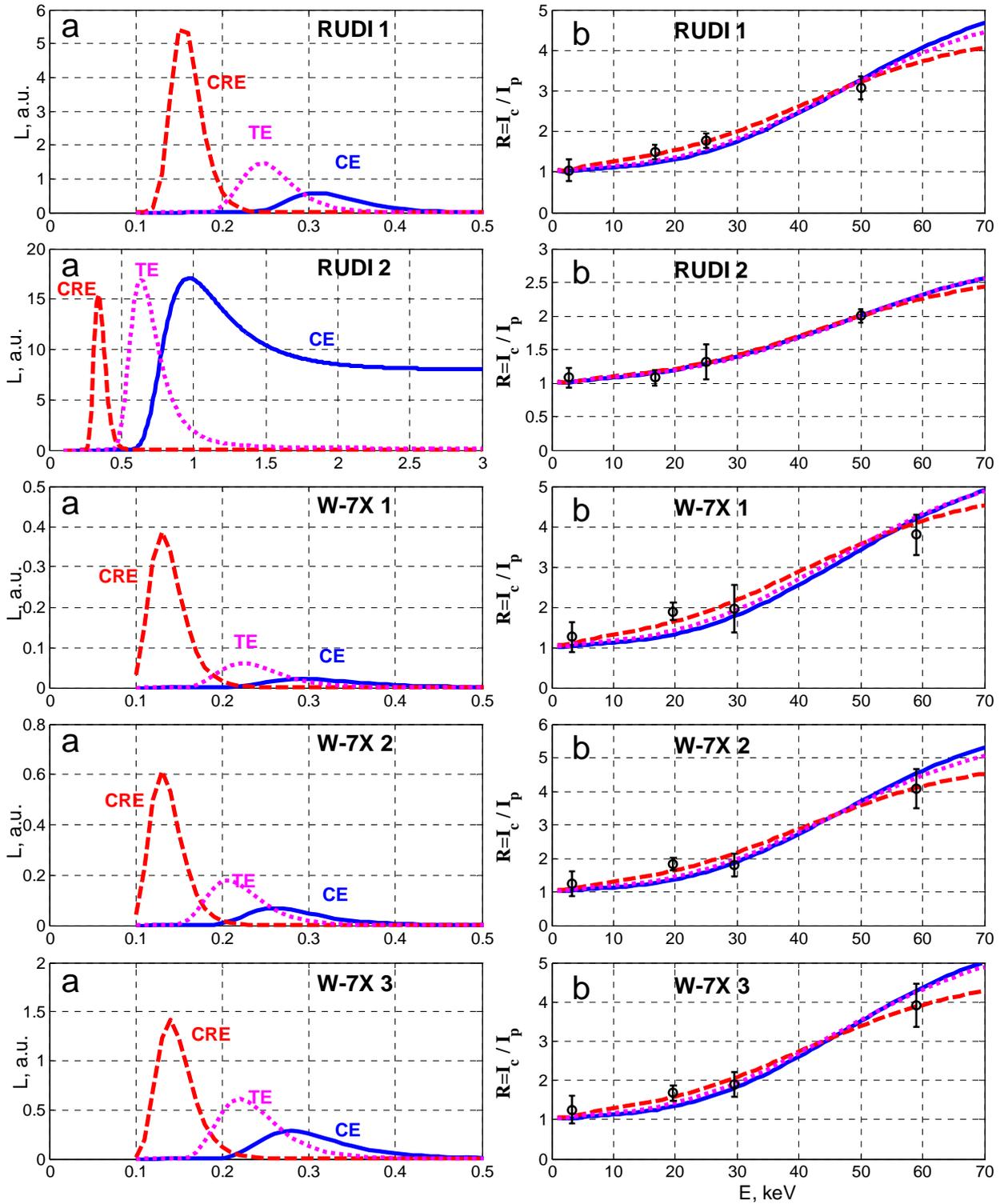

*Fig. 3. Comparison of experimental measurements with model calculations of R value; a – likelihood functions for different models of equilibrium; b – experimental values of R (dots) fitted by calculation with maximal likelihood values of <nl> (lines). Solid line – **CE** model, dashed line – **CRE** model, dotted line – **TE** model. The models are fitted to experimental data with neutralizing target thickness <nl> as free parameter.*

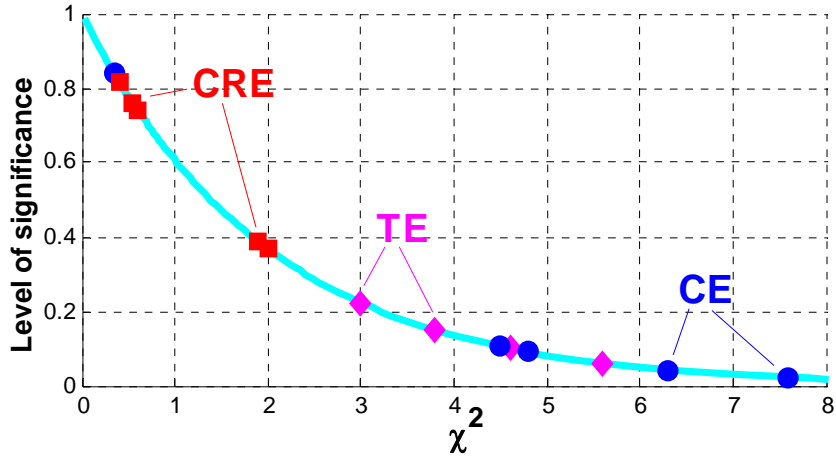

*Fig.4 Levels of significance for consistency of the models (line) with experimental data for different sets of measurements and models of equilibrium (dots), circles- **CE** model, diamonds – **TE** model, squares – **CRE** model*

The data of . $\chi^2$ values from the table 1 together with corresponded level of significance (that is 1-CDF value) is presented on the fig. 4.The $\chi^2$ values for **CE** and **TE** equilibrium models in the four of the five measurements exceed the values accordingly 4.5 and 3.0 that correspond to chi-square cumulative distribution function values 0.89 and 0.78. In the remaining one measurement (RUDI 2) all models predicts practically identical fit of the data. Consequently experimental data pointed out that **CE** and **TE** models of equilibrium should be rejected on the significance level 0.11 and 0.22.

## CONCLUSION

Comparison of available experimental data of intensities of Doppler-shifted $H_\alpha$ radiation of neutral beams with model calculations points out that most probably collision-radiative equilibrium of excited state that correspond to moderate level of l-mixing is realized. Traditional coronal model of sublevel population must be rejected on the significance level 10%. Accordingly effects of l-mixing should be taken into account for interpretation of results of Doppler-shift spectroscopy measurements. Additional special-purpose measurements with independent control of neutralization efficiency would be useful to confirm this result.

## ACKNOWLEDGMENTS

This work was financially supported by the Ministry of Education and Science of the Russian Federation Government, and the programs of fundamental researches of RAS and SB RAS.